\documentclass{elsart}
\usepackage[dvips]{graphicx}
\usepackage{longtable}
\usepackage{cite}
\DeclareGraphicsExtensions{.eps}
\begin{document}

\begin{frontmatter}

\title{Vacancy diffusion in the Cu(001) surface I: An STM study}

\author[KOL]{R. van Gastel\thanksref{corrauth}},
\author[IL] {E. Somfai\thanksref{ellaks.present.address}},
\author[KOL]{S. B. van Albada},
\author[IL]{W. van Saarloos} and
\author[KOL]{J. W. M. Frenken}

\thanks[corrauth]{Corresponding author}
\thanks[ellaks.present.address]{Present address: Department of
Physics,
  University of Warwick, Coventry, CV4~7AL, U.K.}

\address[KOL]{Universiteit Leiden, Kamerlingh Onnes Laboratory, PO Box 9504,
2300 RA Leiden, The Netherlands}
\address[IL]{Universiteit Leiden, Instituut-Lorentz, PO Box 9506,
2300 RA Leiden, The Netherlands}

\begin{abstract}
We have used the indium/copper surface alloy to study the dynamics of surface
vacancies on the Cu(001) surface. Individual indium atoms that are embedded
within the first layer of the crystal, are used as probes to detect the rapid
diffusion of surface vacancies. STM measurements show that these indium atoms
make multi-lattice-spacing jumps separated by long time intervals. Temperature
dependent waiting time distributions show that the creation and diffusion of
thermal vacancies form an Arrhenius type process with individual long jumps
being caused by one vacancy only. The length of the long jumps is shown to
depend on the specific location of the indium atom and is directly
related to the lifetime of vacancies at these sites on the surface. This
observation is used to expose the role of step edges as emitting and
absorbing boundaries for vacancies. 
\end{abstract}

\end{frontmatter}

\section{Introduction}
\label{sec:intro}

Over the past decade the Scanning Tunneling Microscope (STM) \cite{bin82}
has been the predominant instrument that has been employed to
study atomic-scale diffusion processes on surfaces.
Many STM studies of surface diffusion phenomena have now been performed
investigating the mobility of steps \cite{wil99,zha97,kui93,poe92}, islands
\cite{mor95,mor01} and adsorbates \cite{bes96}. The STM has also been
the instrument of choice in the study of adatom diffusion and the role
of adatom diffusion processes in crystal growth \cite{bru98}.
However, the study of the diffusion of \emph{naturally} occurring adatoms and
vacancies is hampered by the finite temporal resolution of the STM.
Adatoms and vacancies both involve two
energy parameters, a formation energy and a diffusion barrier. Typically,
these are such that there is range of temperatures at which either species
is present only in very low numbers, while being extremely mobile, far too
mobile to be imaged with an STM. Lowering the temperature to reduce
the diffusion rate, so that it can be monitored with an STM, also causes the
density of naturally occurring adatoms and vacancies to drop
to extremely low values, so low that no vacancies and adatoms can anymore
be observed. The study of adatom diffusion with Field Ion Microscopy (FIM)
or STM \cite{kel94,bru98} has been made possible though by the simple fact
that adatoms can be deposited on a surface at low temperatures from an
external evaporation source. In the case of FIM, they can also be produced
by means of field evaporation \cite{tso72,tsong90}. No such possibility
exists for surface vacancies and it is for this reason that the role of
monatomic surface vacancies in the atomic-scale dynamics of single
crystal metal surfaces has remained relatively unexposed sofar.

In this paper we present a detailed study of the diffusion of monatomic
vacancies in the first layer of Cu(001). For this purpose we employ indium
atoms that are embedded in the outermost copper layer. We show that the
indium atoms diffuse through the first layer with the assistance of surface
vacancies \cite{gas00,gas01}. The details of the motion of the
indium contain information on the diffusion of the surface
vacancies. The theoretical framework which we use
to interpret our measurements is described in full in the accompanying paper,
which we shall refer to as paper II \cite{somfut}. The diffusion of
vacancies leads to an unusual, concerted type
of motion of surface atoms and causes significant mobility of the
Cu(001) surface as a whole at temperatures as low as room temperature.

\section{Experimental procedures}
\label{sec:exp}

The Cu sample (4.8 $\times$ 4.8 $\times$ 2.0 mm$^3$) was
spark cut from a 5N-purity single crystal ingot. The crystal
was chemically etched and then polished parallel to the
(001)-plane \cite{koper}. Prior to mounting the crystal in the UHV system,
we heated it to 1150 K for 24 hours in an Ar/H$_2$ (20:1)
atmosphere to remove sulphur impurities from the bulk of the crystal.
After introduction into the vacuum chamber, the sample
surface was further cleaned through several tens of cycles of
sputtering with 600 eV Ar$^+$ ions and annealing to 675 K. Initially,
after approximately every fifth cycle the surface was exposed to a
few Langmuir of O$_2$ at a temperature of 550 K to remove carbon
contamination from the surface. The frequency with which the
surface was exposed to oxygen was lowered as the preparation
progressed.

All experiments were performed with the programmable
temperature STM constructed by Hoogeman et al. \cite{hoo98}.
At the start of the experiments, STM images showed a clean,
well-ordered surface with terrace widths up to 8000 \AA.
Small quantities of indium were deposited on the surface
from a Knudsen cell.

\section{Vacancy-mediated surface diffusion}
\label{sec:vacdiff}

Vacancies have been invoked in the past to explain
the incorporation of foreign atoms into a surface \cite{kla88,fin90,ros01}
or the ripening of adatom islands \cite{han97}. The vacancy-mediated
diffusion mechanism of embedded atoms was first proposed for the
motion of Mn atoms in Cu(001) during the formation of a surface
alloy \cite{flo97,flo97a}. Our STM investigation of the diffusion of
indium atoms embedded within the first layer of a Cu(001) surface
was the first to prove unambiguously that this motion takes place with the
help of surface vacancies \cite{gas00,gas01}. More recently, Pd atoms
were shown to diffuse through the Cu(001) surface by the same diffusion
mechanism \cite{gra01}. In this section, we show how we can conclude
that the indium atoms diffuse through the surface with the help of
surface vacancies, without being able to observe them directly.

\begin{figure}
\includegraphics*[width=10.0cm]{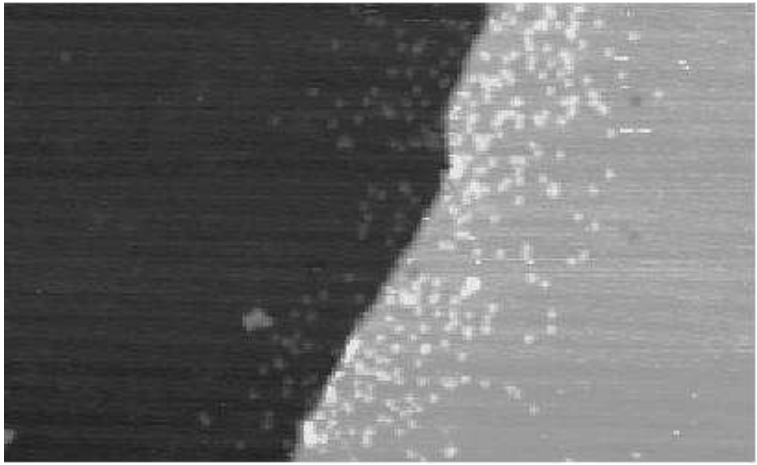}
\caption{662 $\times$ 404 \AA$^2$ STM image of a monatomic height step
on a Cu(001) surface taken 38 minutes after the deposition of 0.03 ML
of indium at room temperature. Embedded indium atoms show up as bright
dots with a height of 0.4 \AA. The image shows a high density of
embedded indium atoms near the step. (V$_t$ = -1.158 V, I$_t$ = 0.1 nA)}
\label{fig:stepview}
\end{figure}

The starting point of our observations is shown in figure \ref{fig:stepview}.
At 38 minutes after the deposition of 0.03 ML of indium at room
temperature, the image shows a high density of indium atoms in the area around
a step edge. Comparison of the apparent height of the indium atoms observed in
this image ($\approx$ 0.4 \AA) with that of indium adatoms that were observed
on the Cu(1~1~17) surface ($\approx$ 2.55 \AA)\cite{gasfut}, shows that the
indium atoms in figure \ref{fig:stepview} are indeed embedded in the first
layer of the crystal. From the image it is obvious that the indium atoms have
been incorporated in the surface through steps: the terraces are not populated
uniformly by indium, the impurities are found only in the direct vicinity of
the steps. This behavior is identical to what has been observed for indium
adatoms on a Cu(1~1~17) surface \cite{gasfut}. Once the indium adatoms have
reached a step and attached themselves to it, they invade the first layer
on both sides of the step and over time slowly infiltrate the
entire first layer of the crystal. At room temperature it takes the
indium atoms typically several hours to spread homogeneously through the
entire surface.

The spreading of the indium atoms through the first layer
implies that they are somehow able to diffuse, whilst remaining embedded
within the surface. The diffusion behavior of the embedded indium atoms
was studied by making series of images of the same area on the copper
surface to form an STM movie of the motion \cite{gasmov}.

\begin{figure}
\includegraphics*[width=10.7cm]{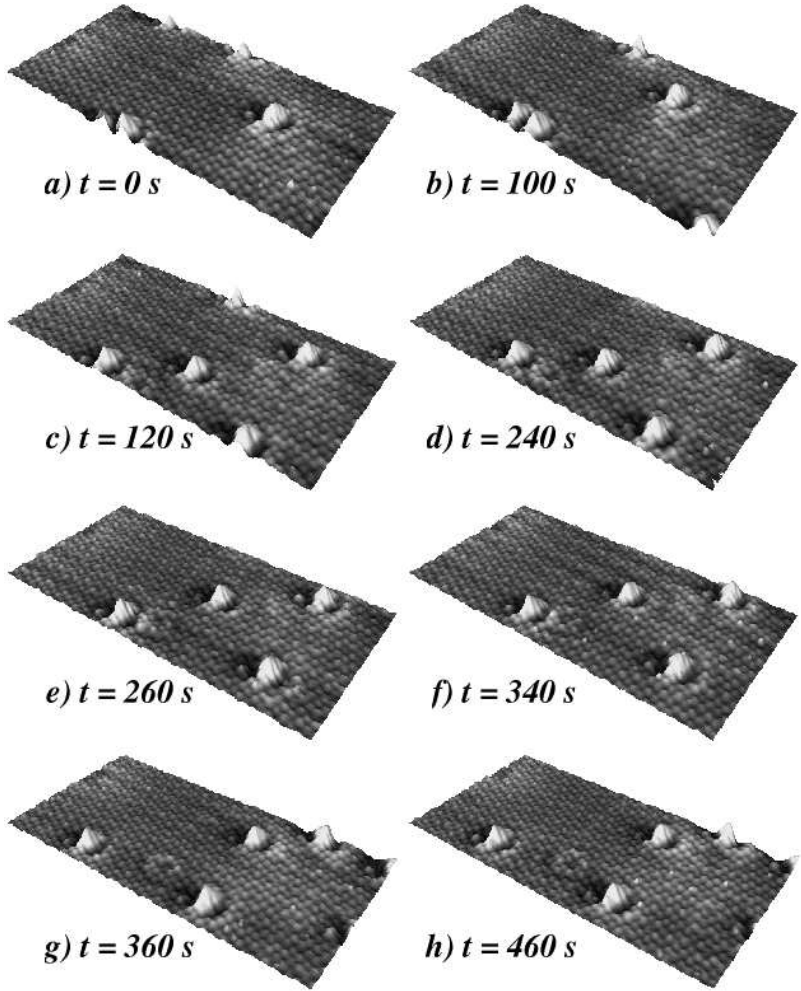}
\caption{140 $\times$ 70 \AA$^2$ STM images of the Cu(001) surface at room
temperature illustrating the diffusion of embedded indium atoms. {\bf (a)}
shows five embedded indium atoms. The right hand panel {\bf (b)} shows that
after 100 s the indium atoms still occupy the same lattice sites. {\bf (c)}
shows the next image in which all indium atoms have made a
multi-lattice-spacing jump. After this jump the indium atoms again stay at
the same lattice site for another two minutes, illustrated by {\bf (d)}.
{\bf (e-h)} show that this pattern of long jumps separated by long time
intervals repeats itself. (V$_t$ = -0.580 V, I$_t$ = 0.9 nA)}
\label{fig:diff3d}
\end{figure}

The diffusion behavior of the indium atoms in these images
is unusual in several respects, as is illustrated by the STM
images in figure \ref{fig:diff3d}.
\begin{itemize}
\item The jumps of the indium atoms are separated by very long time
intervals. At room temperature these intervals can be as long as a
few minutes.
\item If the indium atoms move between two images, they typically move
over several atomic spacings. Na\"{\i}vely one would expect them to
make single, monatomic hops.
\item Nearby indium atoms show a strong tendency to make their jumps
simultaneously. If the indium atoms move independently from one another,
they should not exhibit such concerted motion.
\end{itemize}
To explain the unusual diffusion behavior we invoke the existence of an
assisting particle. First of all,
as the assisting particle is invisible in the STM-images we have to assume
that it is too mobile to be imaged with the STM. Second, if the assisting
particle is present in very low numbers only, this could account for the
long waiting times. Third, a multiple encounter between the assisting particle
and the indium atoms may lead to a jump over several atomic spacings.
And finally, if the assisting particle helps to displace one indium atom,
there is a high probability that nearby indium atoms will also be displaced,
leading to a concerted motion of the indium atoms. It will be shown in section
\ref{sec:quanta} that also the shape of the distribution of jump lengths
forms direct evidence for a mechanism in which the motion is only possible
by virtue of an assisting entity.

We now consider the following possibilities:
\begin{itemize}
\item Diffusion of embedded atoms with assistance of an adsorbed
residual gas molecule.
\item Diffusion of embedded atoms through exchange with copper adatoms.
\item Diffusion of embedded atoms through exchange with surface vacancies.
\end{itemize}

If the diffusion of an embedded indium atom were assisted by an adsorbed
molecule from the residual gas in the vacuum system, the rate of long jumps of
the indium atoms should depend on the rate at which these gas molecules
are adsorbed. The diffusion mechanism could be similar to the one
that was observed for Pt on Pt(110) \cite{hor99}, where enhanced diffusion of
a Pt adatom was enabled by adsorption of a H atom. For the diffusion of the
indium atoms, such a mechanism can however not be active. After desorption of
the residual gas molecule, it is no longer present on the surface and is
therefore not available to assist other indium atoms in making long jumps. The
simultaneous jumps of the indium atoms cannot be explained with this
mechanism. Secondly, as we will see in section \ref{sec:quanta}, the rate of
long jumps depends strongly on the surface temperature. For the adsorption
rate from the residual gas in a UHV system, such a dependence should not be
expected. Third, the length of the long jumps will depend on how long the
residual gas molecule resides at the indium atom. As the residence time of
the molecule goes down exponentially with temperature, the jump length of the
indium atom should do the same. This is definitely not what we observe
(section \ref{sec:quanta}). Diffusion of the indium atoms with the
assistance of adsorbed gas molecules can thus be ruled out.

\begin{figure}
\includegraphics*[width=10.7cm]{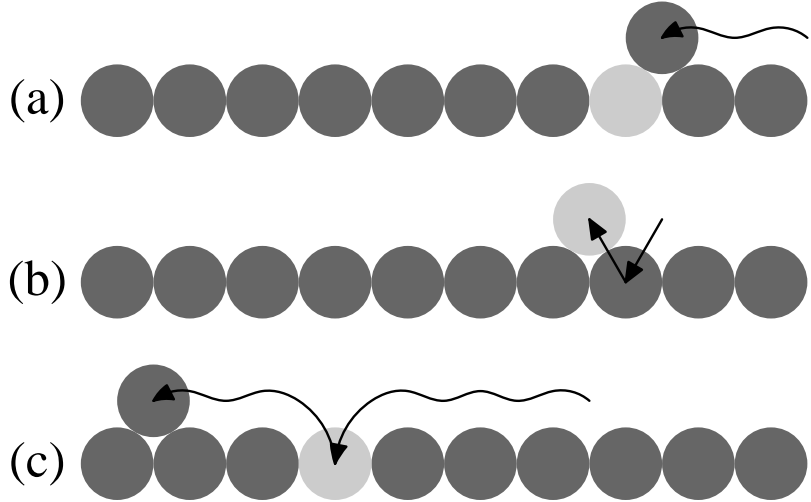}
\caption{A cross-sectional view of an exchange process of indium(bright)
with a copper(dark) adatom, leading to a long jump of the indium atom.
{\bf (a)}, A copper adatom arrives at the embedded indium site, either
through normal hopping or exchange hopping(not shown). {\bf (b)}, The copper
adatom changes places with the indium atom. {\bf (c)}, The indium adatom now
makes one or more hops over the surface before it reinserts itself
into the first layer. A multiple encounter between the copper adatom
and the indium may lead to even larger dispacements.}
\label{fig:cartoonad}
\end{figure}

\begin{figure}
\centering
\includegraphics*[width=10.7cm]{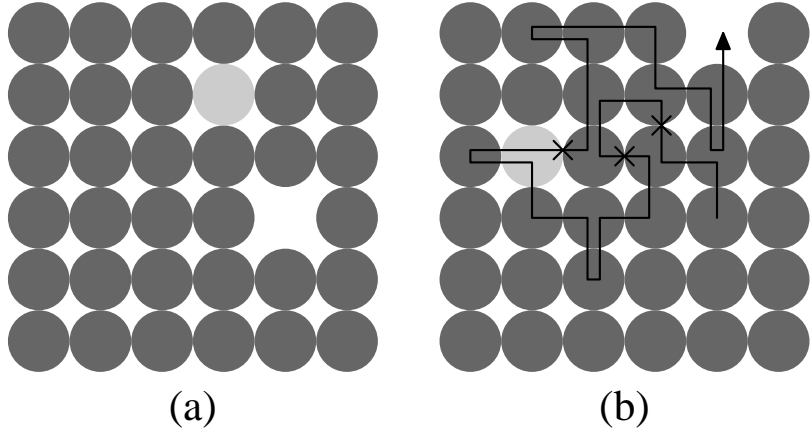}
\caption{A ball-model (top view) of a diffusion event in which the passage
of a surface vacancy
leads to a multi-lattice-spacing displacement of the indium atom(bright).
The arrow indicates the random walk pathway of the vacancy, and the
indium-vacancy exchanges are marked with crosses to show the pathway of
the indium between its beginning and endpoints.}
\label{fig:cartoonvac}
\end{figure}

This leaves the possibility of place exchange with either a copper adatom
or a surface vacancy. For exchange with a copper adatom the diffusion
mechanism is illustrated in figure \ref{fig:cartoonad}.
In the measurements, embedded indium atoms are observed to make jumps
of several atomic spacings. For the adatom mechanism, this would imply that
the indium adatoms reinsert themselves into the terrace after making at
most a few hops. However, as was already seen in low temperature measurements
\cite{gasfut}, after deposition on the Cu(001) surface, indium adatoms do not
insert themselves into the first layer. Instead they perform a random walk on
top of the Cu(001) surface and attach to steps. The same conclusion can be
drawn from the room temperature deposition experiment, in which the indium is
deposited homogeneously onto the surface. The adatom mechanism would lead to
a homogeneous distribution of embedded indium atoms directly after deposition.
By contrast, figure \ref{fig:stepview} clearly shows a high density of
embedded indium atoms near a step shortly after deposition.
The copper adatom exchange mechanism can therefore be ruled out with
confidence. We identify the exchange with surface vacancies as the mechanism
responsible for the observed diffusion of indium through the surface.
Figure \ref{fig:cartoonvac} illustrates the vacancy-mediated diffusion
mechanism.

Having shown that the diffusion of indium in Cu(001) is vacancy-mediated,
we now argue why the vacancies never show up in the STM images, by use
of energies from embedded atom calculations \cite{brecalc}.
Using the formation energy of a surface vacancy in Cu(001) of 517 meV,
we predict a density of vacancies of 1$\cdot$ 10$^{-9}$
at room temperature. Assuming an attempt frequency of 10$^{12}$ Hz, combined
with the calculated diffusion barrier of 0.35 eV, we expect a vacancy jump
rate of 10$^6$ Hz at room temperature. This means that we cannot ``see'' the
vacancies for two reasons. At every instant in time the probability for
even a single vacancy to be present in the STM scan area is very low. The
rate at which each vacancy moves is much higher than the imaging rate of
the STM and it also exceeds the bandwidth of the STM's current preamplifier
by several orders of magnitude. Lowering the temperature to slow the vacancies
down would reduce the vacancy density even further, while increasing the
temperature to create more vacancies would lead to an even higher vacancy
mobility. It is therefore only via the artificial introduction of vacancies
above the equilibrium concentration or through the vacancy-induced diffusion
of tracer particles, such as the embedded indium atoms, that we can detect
the presence and motion of the vacancies.

\section{Quantitative analysis}
\label{sec:quanta}

The vacancy-mediated diffusion mechanism of indium atoms in a Cu(001)
surface provides an unprecedented opportunity to probe the properties
of monatomic surface vacancies. The role of the indium atom is that
of the tracer particle. Since it can be detected with the STM it reveals
precisely when a vacancy passed through the imaged region, and its
displacement provides a measure for how many encounters the indium atom
has had with the vacancy. In this section a quantitative
analysis of the diffusion of embedded indium atoms is presented and is
used to evaluate some of the fundamental energy parameters of surface
vacancies.

\subsection{Jump length distribution}
\label{sec:slp-qa-jld}

\begin{figure}
\centering
\includegraphics*[width=10.0cm]{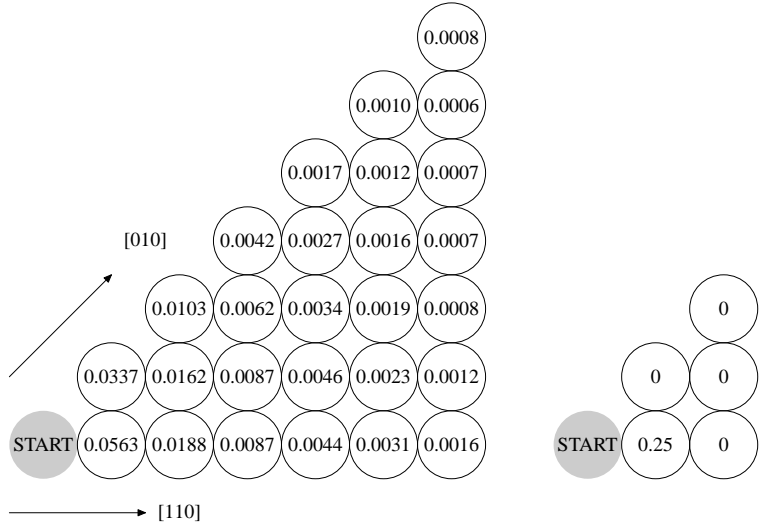}
\caption{The distribution of jump vectors measured from 1461 jumps
that were observed in STM movies at 319.5 K.
Plotted is the probability for jumps of an indium atom from its
starting position to each of the nonequivalent lattice sites shown.
Probabilities have been normalized so that the probabilities for the
entire lattice (not just the non-equivalent sites) add up to one.
In contrast to the calculations in paper II \cite{somfut},
the probability for a jump of length zero cannot be measured with
the STM and has been put to zero. To illustrate the unusual
diffusion behavior, the jump vector distribution for the case of
simple hopping is plotted to the right.}
\label{fig:trijvd}
\end{figure}

The diffusion of indium atoms in the surface proceeds through
multi-lattice-spacing jumps separated by long time intervals.
The multi-lattice-spacing nature of the diffusion is illustrated
by the jump vector distribution which is plotted in figure \ref{fig:trijvd}.
The jump vector distribution shows that there is a significant probability
for the indium atom to jump as far as five or six atomic spacings. In terms
of the vacancy-mediated diffusion mechanism, if the vacancy were making an
ordinary random walk and were not influenced by the presence of the indium,
standard random walk theory \cite{fel68} gives that on average the vacancy
and the indium atom must change places as often as twenty to thirty times to
give such a large displacement.

\begin{figure}
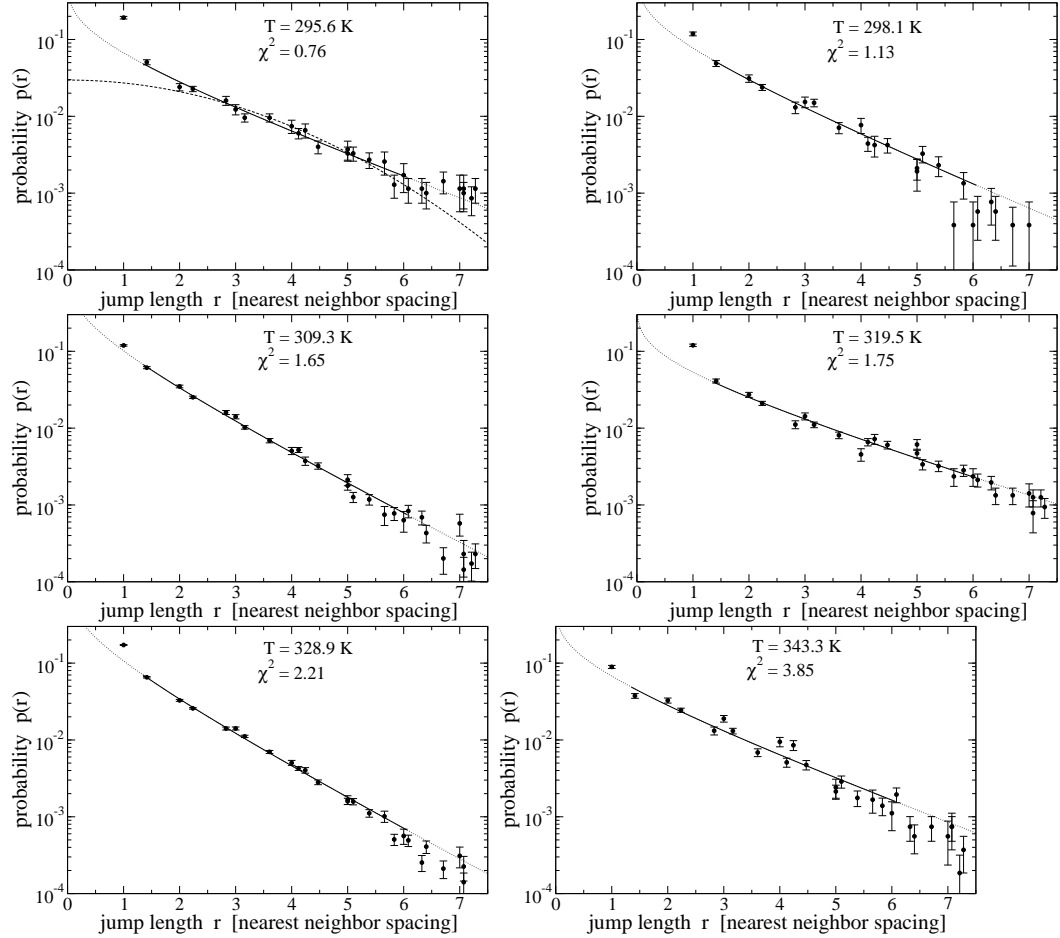

\includegraphics*[width=6.35cm]{t290}
\includegraphics*[width=6.35cm]{t293}
\includegraphics*[width=6.35cm]{t305}
\includegraphics*[width=6.35cm]{t320}
\includegraphics*[width=6.35cm]{t330}
\includegraphics*[width=6.35cm]{t345}
\caption{Radial jump length distribution for embedded indium atoms
measured for six different temperatures. The data points have been fitted
with the modified Bessel function of order zero that is expected for
the vacancy-mediated diffusion mechanism \cite{gas01,somfut}. The fits
were made to the data points for jump lengths from $\sqrt{2}$ to 6, as
indicated with the solid part of the curve in each panel. The normalized
goodness of fit $\chi^2$ for this range is indicated. The dashed curve
in the upper left panel is the best-fit Gaussian curve for the same
data range.}
\label{fig:slp-2djvd}
\end{figure}

In paper II and in ref. \citen{gas01} we have derived that for the
vacancy-mediated diffusion mechanism, one expects the shape of the length
distribution of the long jumps to be that of a modified Bessel function of
order zero. To verify this, the data of figure \ref{fig:trijvd} were
replotted in figure \ref{fig:slp-2djvd} as the radial jump length distribution
function. Figure \ref{fig:slp-2djvd} contains six plots for the distribution
at different temperatures.
Each distribution can be fitted very well
with the modified Bessel function, confirming the vacancy-mediated
diffusion mechanism for the indium atoms. Discrepancies between the
expected and measured probabilities for jumps of unit length are an
artifact of the automated procedure that was used to analyse
the images \cite{jak01,error}. For comparison, one of the panels
contains the best-fit Gaussian curve that should be followed
for ordinary, i.e. non-assisted diffusion. The only free parameter
used in the fits is the recombination probability $\hat p_{rec}$
for vacancies at steps, between subsequent encounters with the same
indium atom. This probability is directly related to the average
terrace width (see the paper II and section
\ref{sec:qamsd}). The terrace widths
that are extracted from the fitting procedure are all within a
factor 2.5 of the ``average'' terrace width of 400 atomic
spacings that was used in the model calculations of paper II.
The variations in this number can be ascribed to the proximity
of steps. The effect of steps will be discussed in more detail in
section \ref{sec:qamsd}.

\subsection{Waiting time distribution}
\label{sec:slp-qa-wt}

\begin{figure}
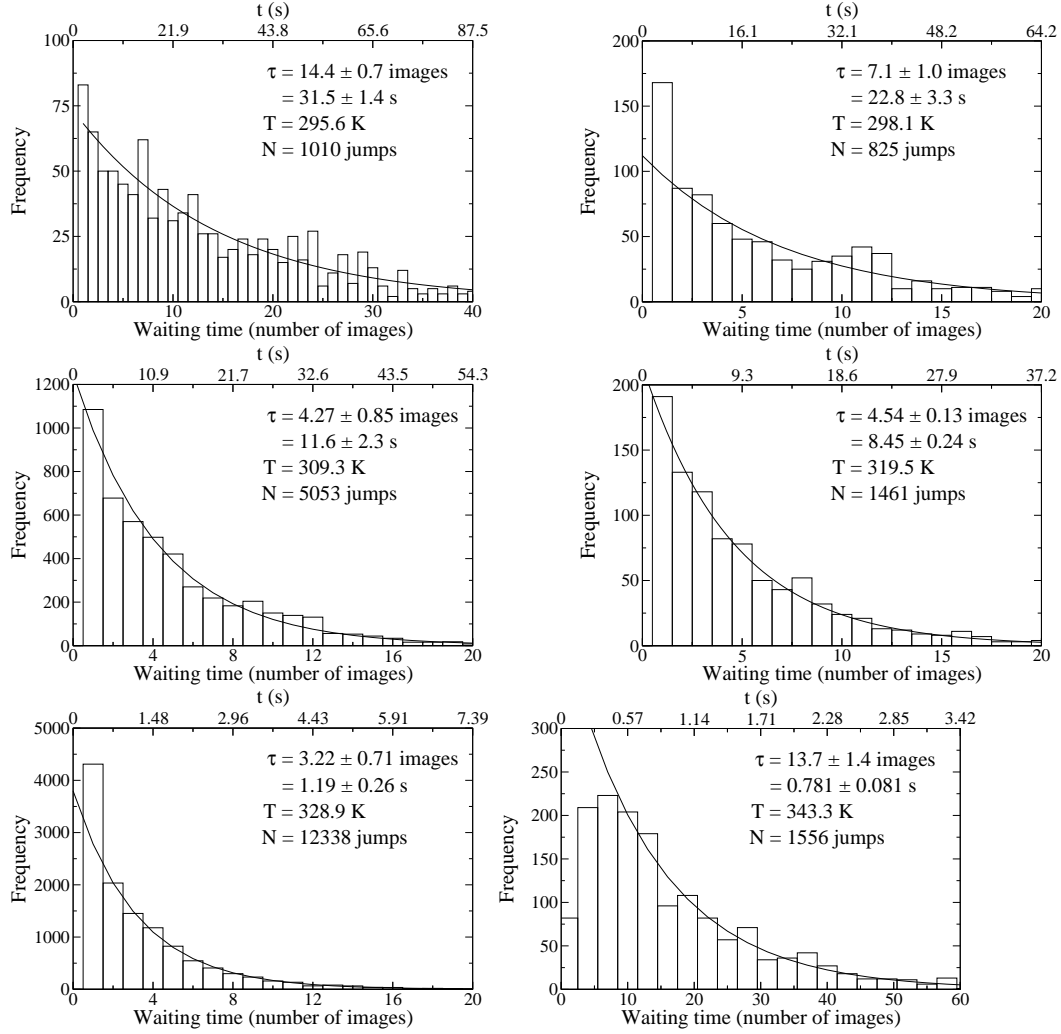

\includegraphics*[width=6.35cm]{wait290k}
\includegraphics*[width=6.35cm]{wait293k}
\includegraphics*[width=6.35cm]{wait305k}
\includegraphics*[width=6.35cm]{wait320k}
\includegraphics*[width=6.35cm]{wait330k}
\includegraphics*[width=6.35cm]{wait345k}
\caption{Waiting time distributions measured for six
different temperatures. All fits are pure
exponentials The time constant $\tau$ is shown in the
graph for each of the distributions. Too high or too low
count rates at short waiting times are an artifact of
the automated analysis scheme of the images \cite{jak01,error}
and are ignored in the fits.}
\label{fig:slp-wtd}
\end{figure}

One of the unusual aspects of the diffusion of embedded
indium atoms is the long waiting time between consecutive
jumps. The distribution of waiting times, expressed here
in number of images, has been plotted for six different
temperatures in figure \ref{fig:slp-wtd}.
As can be seen from the figure, all measured
distributions are purely exponential, with a time constant
$\tau$ that decreases with increasing temperature. The exponential
shape of the distributions shows that the waiting time of an
indium atom to its next long jump is governed by a Poisson process
with rate $\tau^{-1}$. This implies that subsequent long jumps
are independent, which we take as proof that they are caused by
different vacancies. The vacancies are created independently at
random time intervals. The frequency with which
a specific lattice site of the Cu(001) surface is visited
by new vacancies is $\tau^{-1}$. This frequency is such that
our STM can easily keep up with the diffusion process up to a
temperature of 350 K.

\subsection{Temperature dependence of the jump frequency}
\label{sec:slp-qatdep}

\begin{figure}
\includegraphics*[width=10.7cm]{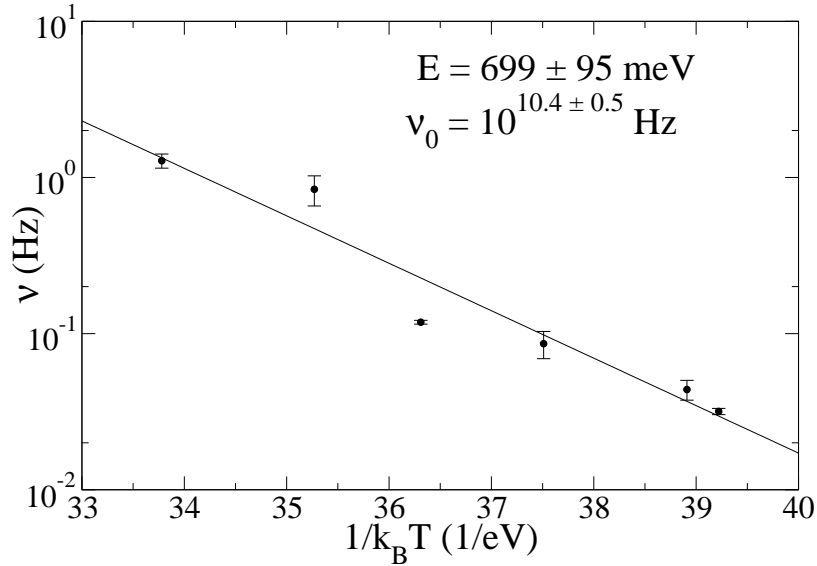}
\caption{Arrhenius plot of the rate of long jumps of the embedded indium
atoms. The activation energy and attempt frequency are shown in the graph
(see also fig. \ref{fig:difftdep})}
\label{fig:arrhenind}
\end{figure}

Simple surface diffusion processes are governed by a single diffusion
barrier $E_D$, which the difffusing species has to overcome to hop from 
one potential well to the next. To verify whether the diffusion of
embedded indium atoms in a Cu(001) surface is indeed a thermally
activated process, the average jump rate
of the embedded indium atoms has been plotted as
a function of $1/k_BT$ in figure \ref{fig:arrhenind}.
The figure shows that the diffusion of embedded indium atoms through
the Cu(001) surface is indeed a thermally activated process with an
activation energy of 699 meV and an attempt frequency of 10$^{10.4}$ Hz.
As will be discussed in section \ref{sec:slp-qaeact}, figure
\ref{fig:arrhenind} does not imply that the process is governed by a single
activation energy.

\subsection{Mean square displacement}
\label{sec:qamsd}

The mean square jump length of the indium atoms is a direct measure for the
number of encounters that an indium atom has on average with a single passing
vacancy. From the six jump length distributions that were shown in section
\ref{sec:slp-qa-jld}, the mean square jump length has been measured. The
results are shown in table \ref{tab:msqdist}.
\begin{table}[b]
\centering
\begin{tabular}{|c|c|} \hline
$T(K)$ & $\langle u^{2} \rangle (latt.spac.^{2})$ \\ \hline
295.6 & 4.37$\pm$0.11 \\ \hline
298.1 & 4.16$\pm$0.13 \\ \hline
309.3 & 3.61$\pm$0.04 \\ \hline
319.5 & 7.67$\pm$0.15 \\ \hline
328.9 & 2.82$\pm$0.02 \\ \hline
343.3 & 5.15$\pm$0.12 \\ \hline
\end{tabular}
\caption{Measured mean square displacement as a function of temperature.}
\label{tab:msqdist}
\end{table}
The values in table \ref{tab:msqdist} exhibit a seemingly random variation
that is much larger than the statistical error margins. As discussed in
paper II \cite{somfut}, we should expect $d^2$ to depend only logarithmically
the terrace width.
Given the fact that all measurements were performed on terraces
with a typical width of a few hundred Angstroms, the considerable variation
in the mean square jump length with temperature is unexpected.

\begin{figure}
\includegraphics*[width=10.7cm]{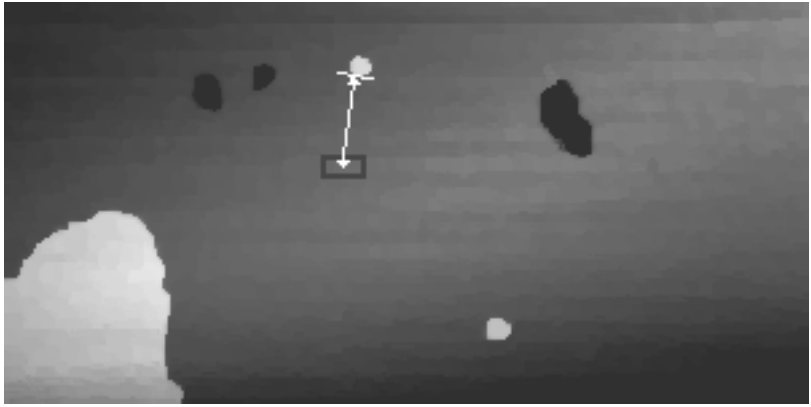}
\caption{2495 $\times$ 1248 \AA$^2$ STM image of the Cu(001)
surface at 343.3 K. The surface area that was selected to perform
the diffusion measurements at this temperature is indicated by the rectangle.
The trajectory to the nearest step has been drawn in the figure. The length
of this trajectory is shown in table \ref{tab:mindist}
(V$_t$ = -1.158 V, I$_t$ = 0.1 nA).}
\label{fig:meassite}
\end{figure}

It is at this point that the precise details of the measurements
become important. After the deposition of the indium the room
temperature measurements were performed and the temperature was then
raised to the values shown in table \ref{tab:msqdist}. Each time after
raising the temperature, the STM was allowed to stabilize its temperature
to the point where the STM images showed no lateral thermal drift. A
suitable area on a terrace was then selected to perform the diffusion
measurements at that specific temperature. No effort was undertaken to
select the same area at all six temperatures. As an example the region that
was selected for the 343.3 K measurements is shown in figure
\ref{fig:meassite}.
From each of the overview scans that were made prior to starting
the diffusion measurements, the distance to
the steps surrounding the measurement site was quantified by
measuring both the logarithm of the minimum distance to a step
as well as the polar average of the logarithm of the distance to
the steps. The measured values are shown in table \ref{tab:mindist}.

\begin{figure}
\includegraphics*[width=10.7cm]{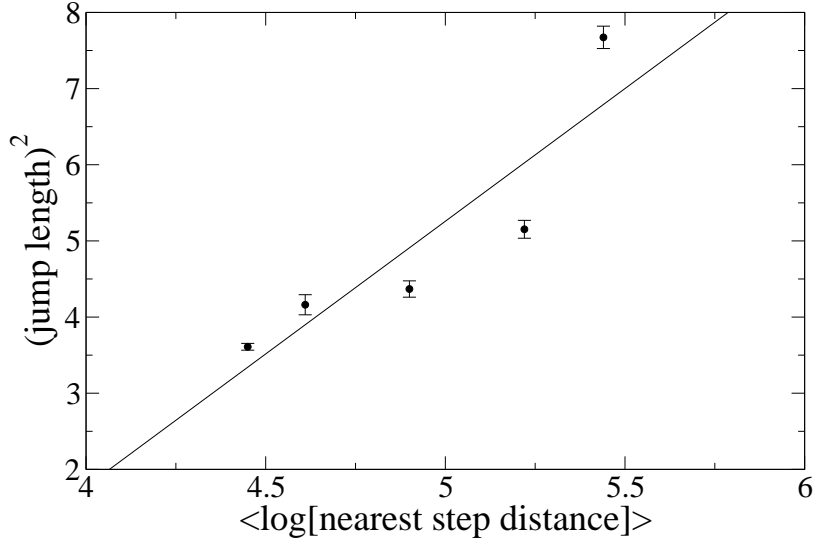}
\caption{Plot of the mean square jump length versus the average
logarithm of the distance of the measurement area to the steps. The graph
shows that measurements that are made further away from a step yield
larger jump lengths. All distances were measured in atomic spacings.}
\label{fig:avdistvslog}
\end{figure}

\begin{table}
\centering
\begin{tabular}{|c|c|c|c|} \hline
$T(K)$ & $\ln (d_{step,min})$ & $\langle \ln (d_{step})\rangle$ &
$\langle u^2 \rangle$ \\ \hline
295.6 & 4.52 & 4.90 & 4.37 \\ \hline
298.1 & 4.46 & 4.61 & 4.16 \\ \hline
309.3 & 4.32 & 4.45 & 3.61 \\ \hline
319.5 & 5.10 & 5.44 & 7.67 \\ \hline
328.9 & 4.48$^*$ & 4.56$^*$ & 2.82 \\ \hline
343.3 & 4.93 & 5.22 & 5.15 \\ \hline
\end{tabular}
\caption{Measured nearest-step-distances as a function of temperature.
All distances have been measured in Cu(001) atomic spacings. The two
values marked with a $^*$ are estimated values: the overview scan at
328.9 K showed only a part of the relevant surroundings of the area
that was used to obtain the jump data.}
\label{tab:mindist}
\end{table}
Both ways to quantify the proximity of steps to the measurement site
show that there is a very clear trend for longer jumps when the steps are
further away. This trend is even more clear from figure \ref{fig:avdistvslog},
where the mean square jump length is plotted versus the average logarithm
of the distance of the measurement site to a step.
In this graph the 328.9 K measurement has not been used as the measurement
area was partly outside the area that was imaged in the overview scan
and the distance to the surrounding steps could not be properly determined.

\begin{figure}
\includegraphics*[width=10.7cm]{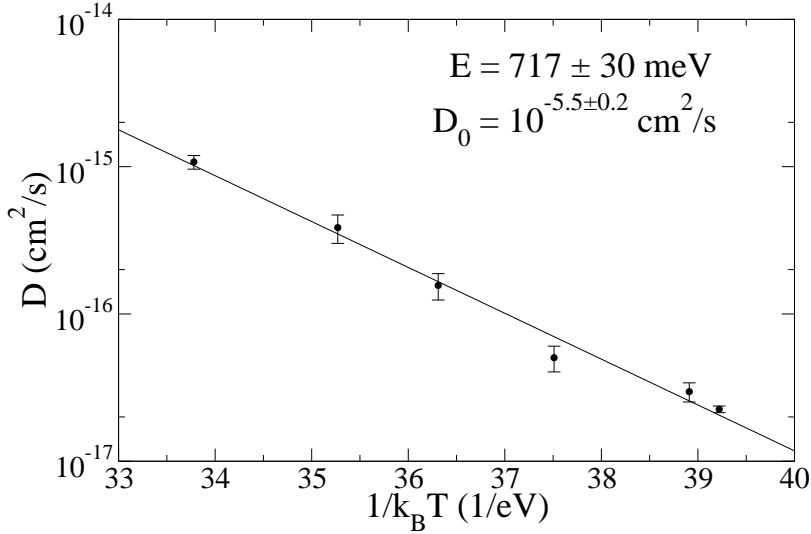}
\caption{Arrhenius plot of the diffusion coefficient D of embedded indium
atoms. The coefficient D was calculated at each temperature by multiplying
the rate of long jumps (figure \ref{fig:arrhenind}) with the mean-square jump
length (table \ref{tab:msqdist}).}
\label{fig:difftdep}
\end{figure}

Because all measurements were performed in thermodynamic equilibrium,
the density of vacancies throughout the terraces is constant.
The observation that the jump length of the indium atoms
depends on the specific geometry in which the measurements
were performed therefore implies that the rate of long jumps is also
position dependent. This in turn means that since all measurements were
performed at different areas, the temperature dependence of the diffusion
of embedded indium atoms can only be properly measured by plotting
the diffusion coefficient D of the indium atoms versus $1/k_BT$
as opposed to the jump rate versus $1/k_BT$ which was plotted in
figure \ref{fig:arrhenind}. This has been done in figure \ref{fig:difftdep}.
We observe that the scatter that was present in figure \ref{fig:arrhenind}
is significantly reduced and a much more accurate value of 717 $\pm$ 30 meV
for the activation energy is obtained. 

\section{A vacancy attachment barrier ?}
\label{sec:slp-schwoeb}

\begin{figure}
\includegraphics*[width=10.7cm]{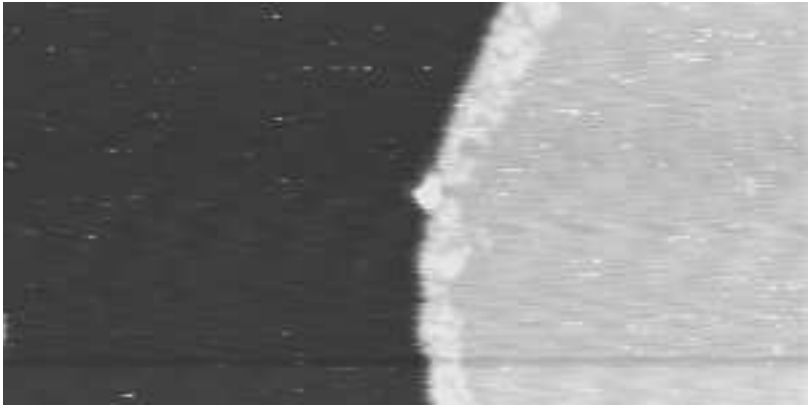}
\caption{830 $\times$ 415 \AA$^2$ STM image of indium being
incorporated into the upper terrace on the right hand side of a step on
the Cu(001) surface. The image was taken at a temperature of 280.4 K during
a slow ramp of the temperature from 130 K, the temperature where
0.015 ML of indium was deposited on clean Cu(001). The indium first
started to be incorporated into the upper terrace at 267 K and
shows up as a light band on the right hand side of the step.
(V$_t$ = -0.672 V, I$_t$ = 0.1 nA)}
\label{fig:uppert}
\end{figure}

Indium atoms that are incorporated into the first layer of the Cu(001)
surface at room temperature via a step appear to have a slight
preference for the upper terrace with respect to the lower terrace.
This effect can be seen from figure \ref{fig:stepview} where approximately
60 \% of the indium atoms have entered the upper terrace and only 40 \%
have gone into the lower terrace. The effect is even more dramatic if
the incorporation of the indium takes place at temperatures below room
temperature. Figure \ref{fig:uppert} shows an image that was obtained at
280.4 K, after indium had been deposited on the surface at 130 K and was
first observed to be incorporated into the surface at a temperature of 270 K.
The image shows that practically all the indium atoms enter the upper
terrace at this temperature. This observation may imply the existence
of a step edge barrier, similar to the Schwoebel barrier for adatoms
\cite{sch66,sch69,ehr66}, but in this case for the attachment of surface
vacancies at the lower side of a step. Using the ratio of the number
of indium atoms that is incorporated into the upper and lower terraces
at 270 K and 300 K, we derive a first estimate of this barrier of 0.8 eV.
The difference in incorporation between the upper and lower terrace
has already been discussed in the context of the Mn/Cu(001) surface
alloy \cite{tre96,iba96,flo97a} and can in principle be attributed to the
difference in the incorporation processes for the upper and lower terrace.
However, the high value of 0.8 eV suggests that other factors play a role
in the preferred incorporation into the upper terrace. One such factor
could be the large amount of indium that is present at the step during
incorporation.

\section{Interpretation of the activation energy}
\label{sec:slp-qaeact}

By definition, the rate at which the indium atom is displaced by a surface
vacancy is the product of the vacancy density at the site next to the indium
atom times the rate at which vacancies exchange with the indium atom. The
activation energy obtained from the temperature dependence of the total
displacement rate will yield the sum of the
vacancy formation energy and the vacancy diffusion barrier. When the
measurements are performed with a finite temporal resolution and if there
is an interaction present between the vacancy and the indium atom, this
simple picture changes. 

\begin{figure}
\includegraphics*[width=10.0cm]{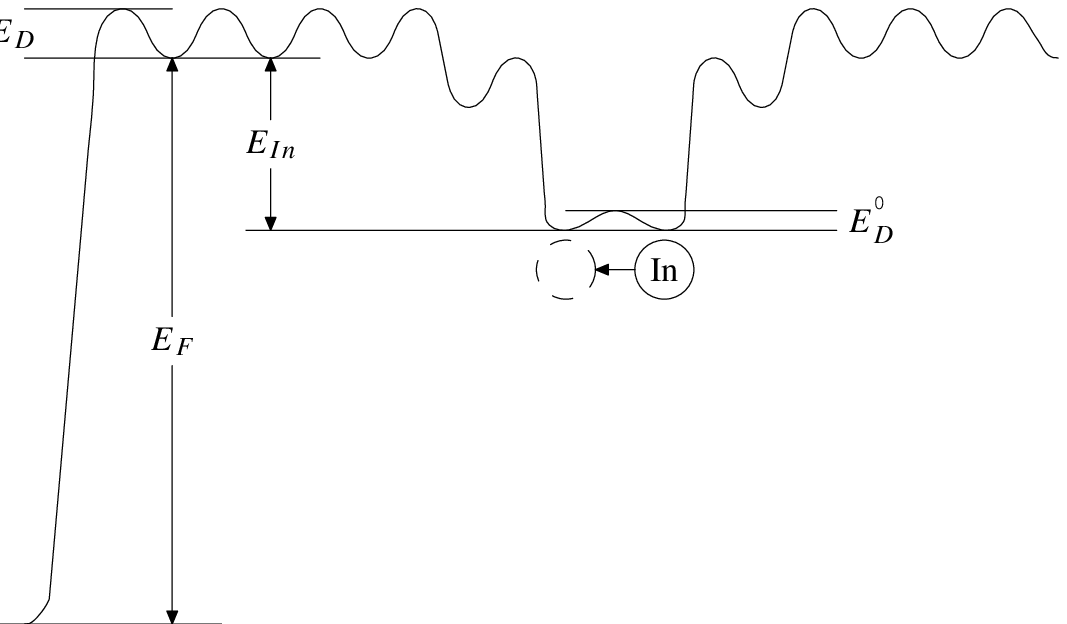}
\caption{One-dimensional energy landscape for a vacancy in the Cu(001) surface.
For this one-dimensional situation the vacancy is formed at the step on the
left and approaches the indium from there. It is only able to displace the
indium atom to the left by one atomic spacing. $E_{F}$ is the formation
energy of a Cu(001)-vacancy, $E_{D}$ is the diffusion barrier for a
Cu(001)-vacancy, $E^{'}_{D}$ is the vacancy-indium exchange barrier and
$E_{In}$ is the binding energy for an indium-vacancy pair.}
\label{fig:energy}
\end{figure}

\begin{figure}
\includegraphics*[width=12.0cm]{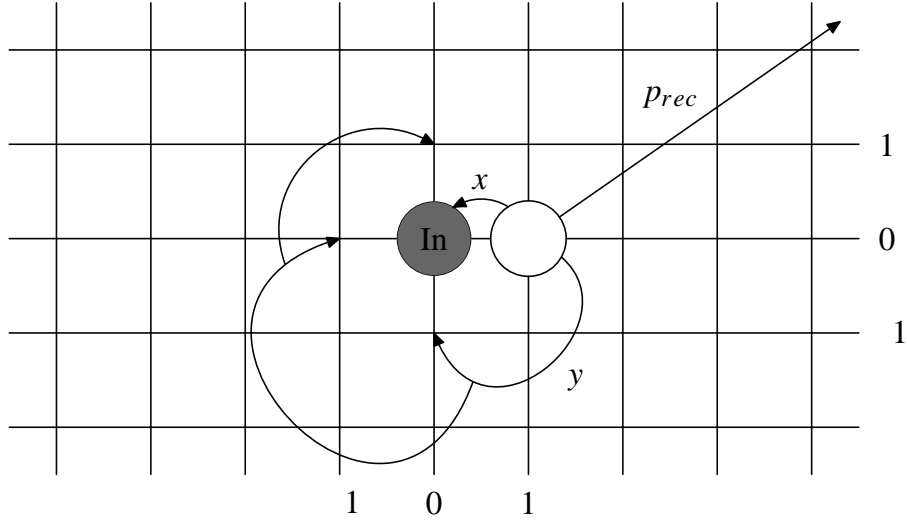}
\caption{The random walk starting situation, with the vacancy directly
next to the indium atom. The probabilities $x$, $y$ and $p_{rec}$ are
introduced in the text.}
\label{fig:start}
\end{figure}

The energy landscape for the vacancy is sketched in figure \ref{fig:energy}.
We now consider the situation where the diffusion is observed with
a finite time resolution. We assume that the time resolution of the
STM-measurements is sufficient to resolve the effect of the entire
random walk of one individual vacancy from that of the next, which
was already shown to be the case
in section \ref{sec:slp-qa-wt}. The starting situation is shown in
figure \ref{fig:start}: an indium atom is embedded in the origin
of a square lattice with a vacancy sitting next to it at (1,0). The only
diffusion barrier that has been modified by the indium atom is the one
which is associated with the vacancy-indium exchange. This is equivalent
to putting $E_{In}=0$ in figure \ref{fig:energy}.

We now define the following additional quantities:

\begin{longtable}[c]{lp{10.0cm}}
$x$ & the probability that in the first step the vacancy
immediately changes places with the indium. \\
$y$ & the probability that the vacancy does not immediately
change places with the indium but that it returns to
any of the four sites neighbouring the indium. \\
$p_{rec}$ & the recombination probability of the vacancy,
i.e. the probability that the vacancy recombines at a
step without ever changing places with the indium. This
probability is determined by the distances to the
nearby steps. \\
\end{longtable}

Note that $x + y + p_{rec} \neq 1$, because, as is indicated in figure
\ref{fig:start}, there is the possibility that the vacancy returns one or more
times to the sites neighbouring the indium atom and then recombines, without
ever exchanging with the indium. Only if one were to define the recombination
probability as $p'_{rec}$, the probability that the vacancy recombines without
first returning to any of the sites neighbouring the indium atom, would
$x + y + p'_{rec} = 1$. The rate at which jumps of the indium are registered
by the STM is determined by the rate at which total displacements with a
length of at least one atomic spacing occur. Given the starting situation of
figure \ref{fig:start}, the probability that the indium is displaced by the
vacancy is given by
\begin{equation}
p_{dis} = 1-p_{rec}=x+y\cdot x + y^{2}\cdot x + ... = \sum_{N=0}^{\infty}y^{N}
\cdot x = \frac{x}{1-y}
\label{eqn:least}
\end{equation}
\newpage
Here the possibility that multiple displacements of the indium may add up to a
zero net displacement is ignored\footnote{This is justified by numerical
calculations presented in paper II where we address this problem for the case
of indium. Our calculations show that the fraction of jumps that have a zero
net length changes only marginally with temperature.}. The precise value of
$x$, the probability to make a successful exchange, is determined by the energy
difference between the modified and unmodified vacancy exchange barrier,
$E_{D} - E^{'}_{D}$.
\begin{equation}
x=x(T)=\frac{\nu_{0}e^{-\frac{E^{'}_{D}}{k_{B}T}}}
       {\nu_{0}e^{-\frac{E^{'}_{D}}{k_{B}T}}+3\nu_{0}e^{-\frac{E_{D}}{k_{B}T}}}
      =\frac{1}{1+3 e^{-\frac{E_{D}-E^{'}_{D}}{k_{B}T}}}
\label{eqn:xoft}
\end{equation}
Upon examining the random walk of the vacancy when it has not exchanged with
the indium but has returned to any of the sites neighbouring the indium, we
see that only the first step of this random walk pathway has a temperature
dependent probability associated with it and $y$ can therefore be split up in
a temperature dependent and a temperature independent part
\begin{equation}
y=y(T)=(1-x(T)) C
\label{eqn:yoft}
\end{equation}
where $0\leq C \leq 1$ is a constant which is determined by the geometry
of the lattice, and especially by the distribution of distances between
the starting position of the vacancy and the absorbing boundaries (steps).
It can be evaluated numerically. Substituting equation (\ref{eqn:yoft})
in (\ref{eqn:least}), the probability to have at least one displacement
is equal to
\begin{eqnarray}
p_{dis} & = & \frac{x(T)}{1-y(T)}\nonumber \\
& = & \frac{x(T)}{1-(1-x(T))\cdot C}\nonumber \\
& = & \frac{x(T)}{(1-C)+Cx(T)}\nonumber \\
& = & \frac{1}{1+3(1-C)e^{-\frac{E_{D}-E^{'}_{D}}{k_{B}T}}}
\label{eqn:least2}
\end{eqnarray}
From equation (\ref{eqn:least2}) it is clear that the final rate of long jumps
will contain exponential terms not only in the numerator, but also in the
denominator. The rate of long jumps should not show normal thermally activated
Arrhenius behaviour.

The observed rate of long jumps is equal to the equilibrium rate
at which vacancies exchange with the indium atom, divided by the average
number of elementary displacements caused by a single vacancy, given that
the vacancy has displaced the indium atom at least once. This average number
of displacements $\langle n \rangle$ is given by
\begin{eqnarray}
\langle n \rangle & = & 1+0\cdot p_{rec}+1\cdot (1-p_{rec})p_{rec}+2\cdot
        (1-p_{rec})^{2}p_{rec}+... \nonumber \\
    & = & \frac{1}{p_{rec}} \nonumber \\
    & = & \frac{1}{1-C} \left( \frac{1}{1-x(T)}-C \right)
\label{eqn:avdisp}
\end{eqnarray}
Using equation (\ref{eqn:avdisp}), the observed rate of long jumps is equal to
\begin{equation}
\nu_{LJ}=\nu_{0}\frac{e^{-\frac{E_{F}+E_{D}+\Delta}{k_{B}T}}}
    {1+\frac{e^{-\frac{\Delta}{k_{B}T}}}{C'}}
\label{eqn:longjumprate}
\end{equation} 
where

\begin{tabular}{ll}
$C'$&$=3(1-C)$\\
$\Delta$&$=E^{'}_{D}-E_{D}$
\end{tabular}

In the case that indium is ``identical'' to copper, so that $\Delta$ = 0,
$\langle$ n $\rangle$ becomes $\frac{4-3C}{3(1-C)}$ and the rate of long jumps
reduces to
\begin{equation}
\nu_{LJ}=\nu_0\frac{e^{-\frac{E_F+E_D}{k_BT}}}{1+\frac{1}{C'}}=
\frac{\nu}{1+\frac{1}{C'}}
\label{eqn:nointer}
\end{equation}
where $\nu$ is the jump rate of atoms in the clean Cu(001) surface.
The denominator corresponds to the number of exchanges contributing to a long
jump.

In the extremely repulsive case $\frac{\Delta}{k_{B}T}\gg -1$ and the average
number of displacements $\langle n \rangle$ becomes approximately equal to $1$.
Equation (\ref{eqn:longjumprate}) reduces to
\begin{equation}
\nu_{LJ}=\nu_{0}e^{-\frac{E_{F}+E_{D}+\Delta}{k_{B}T}}
   =\nu_{0}e^{-\frac{E_{F}+E^{'}_{D}}{k_{B}T}}
\label{eqn:extremerep}
\end{equation}
In the extremely attractive case $\frac{\Delta}{k_{B}T}\ll 1$ and the average
number of jumps $\langle n \rangle$ is $\frac{1}{1-C}\cdot
\frac{1}{3}e^{-\frac{\Delta}{k_{B}T}}$. Equation (\ref{eqn:longjumprate}) now
reduces to
\begin{equation}
\nu_{LJ}=3\nu_{0}(1-C)e^{-\frac{E_{F}+E_{D}}{k_{B}T}}
\label{eqn:extremeatt}
\end{equation}
We immediately see the fundamental difference between attraction and
repulsion. The apparent activation energy for strong attraction is
identical to that for the copper surface itself, while it is larger
in the case of repulsion. This asymmetry between attraction and repulsion
is caused by the fact that for moderate to strong attraction, the probability
$x$ rapidly approaches unity, meaning that the arrival of a vacancy next to
the indium is almost guaranteed to cause a long jump. In contrast, in the
case of repulsion, the probability $x$ scales with the Boltzmann-factor
containing the exchange barrier of the vacancy and the embedded atom.

In general, the Arrhenius plot of the log of the
rate of long jumps versus $\frac{1}{k_{B}T}$ is nonlinear:
\begin{equation}
\ln{\left( \nu_{LJ} \right)}=\ln{\left( \nu_{0} \right)}
	-(E_{F}+E_{D}+\Delta)\frac{1}{k_{B}T}
        -\ln{\left( 1+\frac{e^{-\frac{\Delta}{k_{B}T}}}{C'} \right)}
\label{eqn:logjumprate}
\end{equation}
The last term on the right describes the departure from ideal Arrhenius
behaviour. With this term expanded to second order in $\frac{1}{k_BT}$, the
rate of long jumps can be rewritten as
\begin{eqnarray}
\ln{\left( \nu_{LJ} \right) } & = &
\ln{\left( \nu_{0}\frac{C'}{1+C'} \right)}+\left( -E_{F}-E_{D}
                +\left( \frac{1}{1+C'}-1\right) \Delta\right)
             \frac{1}{k_{B}T} \nonumber \\
       &   & -\frac{C'\Delta^{2}}{2(1+C')^{2}}\frac{1}{(k_{B}T)^{2}}+...
\label{eqn:finallograte}
\end{eqnarray}
For practical values of
$\frac{1}{k_{B}T}$ we are in the limit $\frac{\Delta}{k_{B}T}\ll-1$
and the measured activation energy is equal to the sum of the vacancy
formation energy, $E_F$, and the vacancy diffusion barrier, $E_D$.

We mention again that the fact that multiple displacements of the indium may
actually add up to zero is ignored. The numerical calculations in paper II
\cite{somfut} show that for indium approximately 8.5 \% of all long jumps
have a net displacement of zero atomic spacings. This fraction varies over
the investigated temperature range by an insignificant amount and will
therefore only result in a small modification of the prefactor in the
temperature dependence.

\section{Discussion}
\label{sec:slp-disc}

The jump length distributions mentioned in this work
all have the shape that is expected for the particle-assisted
diffusion mechanism. All distributions can be accurately fitted
with the modified Bessel function, reconfirming that the diffusion
of indium through the first layer takes place with the aid of
some particle, which we deduced  to be a vacancy by looking
in detail at the incorporation process of the indium atoms.
Having used the jump length distributions to validate the model described in
the accompanying paper, we can now use this model to calculate
the diffusion coefficient of copper atoms in a clean Cu(001) surface.
This involves no more than taking the interaction between the
vacancy and the indium atom out of the model such that the vacancy
performs an unbiased random walk. At room temperature the
average jump length of copper atoms in the clean Cu(001) surface
is calculated to be 1.6 atomic spacings. Using the rate of long jumps of
the indium atoms that was measured at room temperature we find that the
diffusion coefficient of surface atoms in the Cu(001) surface
as a result of vacancy mediated diffusion is 0.42 \AA$^{2}\cdot$s$^{-1}$.  
Using the attempt frequency that was obtained from figure \ref{fig:arrhenind},
we find that at room temperature every site on the surface is visited by a
new vacancy on average once every 32 s.

The sum of the Cu(001) surface vacancy formation and migration energy
is equal to 717 $\pm$ 30 meV. The sum of the EAM calculated diffusion barrier
(0.35 eV) and the formation energy (0.52 eV) amounts to 0.86 eV. The fact that
this result is too large is not entirely unexpected as all diffusion barriers
that were calculated with EAM were too large when compared with experiments.
However, the value of 0.35 eV was already obtained by dividing the true EAM
value by a factor 1.7 \cite{brecalc}. Using the measured sum and the
calculated EAM formation energy of 0.52 eV, we find a lower
value of 0.20 eV for the diffusion barrier of a Cu(001) surface
vacancy. Using the vacancy formation energy of 0.485 eV from
\cite{gra01} that was obtained through first-principles calculations we obtain
a vacancy diffusion barrier of 0.232 eV. With such a low diffusion barrier, we
expect a room temperature vacancy diffusion rate of $10^8$ Hz.

The interpretation of the activation energy as the sum of the vacancy
diffusion barrier and the vacancy formation energy is supported by
measurements of vacancy-mediated diffusion of Pd in Cu(001) \cite{gra01},
which reveal a value of 0.88 $\pm$ 0.03 eV for the sum of the vacancy formation
energy and the Pd-vacancy exchange barrier. The higher activation
energy is a direct consequence of the fact that vacancies and embedded
palladium atoms repel (see section \ref{sec:slp-qaeact}).  To be able to
experimentally separate out the two energy parameters, the vacancy formation
energy and the vacancy diffusion barrier, independent measurements are needed
of one of the two energies. One possibility is to follow artificially created
surface vacancies at low temperature \cite{kob93,mol98,may01}. Such a
measurement would yield the diffusion barrier for a monatomic surface vacancy
which could then be combined with the 717 meV that we measured to obtain the
vacancy formation energy.

The variation in mean square jump length that was observed with the
distance to the nearest step confirms the
idea that surface vacancies are formed and annihilated at steps.
The creation of a surface vacancy through the expulsion of an atom
from a section of flat Cu(001) terrace has been predicted to cost 984 meV
\cite{sto94}. Because of this high energy for the creation of an adatom-vacancy
pair, at most temperatures surface vacancies are formed almost exclusively at
steps. This means that the steps section the surface into separated areas
through which the vacancies are allowed to diffuse. The vacancies are
annihilated when they reach a step. A vacancy which exchanges with an
indium atom near a step has a relatively high probability to recombine at
the step as opposed to making another exchange with the indium atom. Indium
atoms near a step will therefore on average make shorter jumps than indium
atoms that are far away from a step. Given a constant density of surface
vacancies throughout the terrace (the measurements were performed in
thermal equilibrium), this implies that the rate of long jumps near steps
will be higher than in the middle of a terrace. Hence, the indium will perform
relatively many jumps of a short length near a step, whereas in the
middle of a wide terrace, indium atoms will tend to jump less frequently,
but given the increased lifetime of the vacancy, they will jump over
longer distances. This observation opens the way to detailed studies of
vacancy creation and annihilation near steps and kinks.

The existence of the vacancy attachment barrier that was proposed
in section \ref{sec:slp-schwoeb} needs to be investigated in more detail.
The influence of the considerable amount of indium that was present at
the step makes it hard to draw detailed conclusions about this barrier.
Further experiments investigating the existence and magnitude of this barrier
at a clean step should be performed with low densities of embedded guest
atoms, e.g. in adatom or vacancy islands. The lifetime of a surface vacancy
in such islands will be affected strongly by the presence of this barrier.
If present, the barrier will lead to significant differences in the
mean square jump length and frequency of long jumps of indium atoms embedded
in equal-size adatom and vacancy islands.

\section{Conclusions}
\label{sec:slp-conc}

Surface vacancies were shown to be responsible for the motion of
embedded indium atoms in the Cu(001) surface. The density of surface vacancies
at room temperature is extremely low, but vacancies are able to diffuse through
the surface at an extremely high rate. In the STM-measurements the
rapid diffusion of these vacancies leads to simultaneous
long jumps of embedded indium atoms. Measurements of the jump length
and jump rate were performed at six different temperatures.
These measurements show that the vacancy-mediated diffusion process
can be accurately described with the model that is presented
in paper II \cite{somfut} and ref. \citen{gas01}, provided that the interaction
between the embedded indium and the surface vacancies is properly taken into
account. The role of steps as sources
and sinks for surface vacancies has been confirmed by measuring
the position dependent jump rate and jump length of the indium
atoms. Near a step the jump rate of the indium
atoms is increased, but at the same time, given the shorter lifetime
of vacancies near a step, the jump length is decreased. The sum of
the vacancy formation and migration energy was measured to be
$717 \pm 30$ meV.

The most far-reaching conclusion concerns
the mobility of the Cu(001) surface itself. The diffusing
vacancy leaves behind a trail of displaced surface copper atoms and thereby
induces a significant mobility within
the first layer of the surface, already at room temperature \cite{gas00}. The
mobility of atoms in the clean Cu(001) surface was evaluated by
applying the model of ref. \citen{gas01} and paper II \cite{somfut} with the
interaction between the tracer particle and the vacancy
ignored. Through this procedure the diffusion coefficient
of Cu(001) surface atoms was measured to be 0.42 \AA$^2\cdot$s$^{-1}$ at
room temperature.

\begin{ack}
We gratefully acknowledge B. Poelsema for help with the preparation of the
Cu-crystal. We acknowledge L. Niesen and M. Ro\c{s}u for valuable
discussions. This work is part of the research program of the ``Stichting
voor Fundamenteel Onderzoek der Materie (FOM),'' which is financially
supported by the ``Nederlandse Organisatie voor Wetenschappelijk
Onderzoek (NWO).''
\end{ack}


\begin{thebibliography}{00}
\bibitem{brecalc}
EAM energies were calculated
using the computer code of M. Breeman, Rijksuniversiteit Groningen. The
absolute value of the diffusion barriers was divided by 1.7
to accomodate the consistent overestimation of the diffusion barriers
in these calculations.
\bibitem{error}
Noise in the STM images will sometimes cause the detected center of mass
of an indium atom to shift by more than one atomic spacing, thereby triggering
a monatomic length jump at too short a time interval.
\end{thebibliography}
\end{document}